%% file: C5.tex
\documentclass[aps,floatfix,twocolumn,prd,nofootinbib,superscriptaddress]{revtex4-1}
\usepackage{graphicx}
\usepackage{color}
\usepackage{psfrag}
\usepackage{dcolumn}
\usepackage{bm}
\usepackage{float}
\usepackage{amssymb}
\usepackage{amsmath}
\usepackage{amsfonts}
\usepackage{longtable}
\usepackage{xspace}
\usepackage{mathrsfs}
\usepackage[export]{adjustbox}

\newcommand{\beq}{\begin{equation}}
\newcommand{\eeq}{\end{equation}}

\newcommand{\be}{\begin{equation}}
\newcommand{\ee}{\end{equation}}
\newcommand{\bea}{\begin{eqnarray}}
\newcommand{\eea}{\end{eqnarray}}
\newcommand{\bes}{\begin{subequations}}
\newcommand{\ees}{\end{subequations}}

\newcommand{\lnL}{\ln \mathcal{L}}
\usepackage{bm}

\usepackage{xcolor}

\begin{document}

\title{The Fifth RIT Catalog of Binary Black Hole Simulations:\\
  Multiple-Resolution Studies of Eccentric Orbits}

\author{Giuseppe Ficarra}
\affiliation{Dipartimento di Fisica, Universit\`a della Calabria, Arcavacata di Rende (CS), 87036, Italy}
\author{James Healy}
\affiliation{Center for Computational Relativity and Gravitation, Rochester Institute of Technology, Rochester, New York 14623}
\author{Carlos O. Lousto}
\affiliation{Center for Computational Relativity and Gravitation, Rochester Institute of Technology, Rochester, New York 14623}

\date{\today}

\begin{abstract}
This fifth release of the RIT public catalog of numerical relativity binary black
hole waveforms (\url{http://ccrg.rit.edu/~RITCatalog}) introduces an additional 248
configurations, prioritizing 197 newly simulated eccentric orbits. This update brings
the catalog to a total of 2129 cases. All waveforms are corrected for center-of-mass
drift and extrapolated to future null infinity. To rigorously estimate waveform errors,
we conduct multiple-resolution convergence studies on 10 eccentric simulations
(up to 33 orbits to merger) using three global resolutions increasing by factors of 1.2,
plus a comprehensive six-resolution study for a single 18-orbit configuration.
We evaluate waveform accuracy by computing mismatches against theoretical
infinite-resolution extrapolations. Additionally, we analyze the convergence properties
of key physical observables: merger times, number of orbits, final masses, final spins,
recoil velocities, and the peak amplitude, frequency, and luminosity of the
gravitational radiation.
\end{abstract}

 \maketitle

\section{Introduction}
\label{sec:intro}

The detection of gravitational waves from binary black holes (BBH) by the LIGO-Virgo-KAGRA (LVK) collaboration~\cite{LIGOScientific:2016aoc, LIGOScientific:2018mvr, LIGOScientific:2020ibl, LIGOScientific:2021djp} has fundamentally transformed our understanding of the universe. Critical to this success is the interplay between observational data and theoretical waveform modeling.
Since the 2005 breakthrough in numerical relativity, which allowed for the full nonlinear simulation of black hole binaries, the modeling of these systems has advanced from the first stable orbits to the systematic population of large-scale waveform catalogs.

The numerical relativity group at the Rochester Institute of Technology (RIT) played a pivotal role in this development using the ``moving puncture'' formalism~\cite{Campanelli:2005dd}—a robust method without excision, that utilizes a specific gauge choice (the $1+\log$ slicing and $\Gamma$-driver shift) to allow black hole singularities to traverse the computational domain freely. This methodology proved robust and scalable, enabling the exploration of the full BBH parameter space.
Thus the group pioneered the exploration of extreme dynamical phenomena. This included the discovery of the ``orbital hang-up'' effect~\cite{Campanelli:2006uy} and the identification of large gravitational recoil velocities (``superkicks'') in spinning binaries~\cite{Campanelli:2007ew, Campanelli:2007cga}, which have profound implications for black hole retention in galactic hosts.

As the field matured towards the detection era, the focus shifted to the systematic mapping of the parameter space to facilitate remnant modeling. By performing hundreds of simulations with varying mass ratios and spin configurations, RIT developed highly accurate phenomenological formulas for the final mass, spin, and recoil velocity of the merger remnant~\cite{Healy:2014yta,Healy:2016lce}. These algebraic expressions allow for inferring the properties of the final black hole from the initial binary parameters in gravitational wave data analysis.

More recently, the group has focused on the production of open public waveform catalogs to support the community. The release of the first RIT Catalog~\cite{Healy:2017psd}, followed by subsequent expansions~\cite{Healy:2019jyf, Healy:2020vre, Healy:2022wdn}, has provided hundreds of aligned and precessing waveforms. These efforts have also pushed the boundaries of numerical relativity into extreme regimes, successfully simulating binaries with near-maximal spins and high mass ratios~\cite{Healy:2018swt, Lousto:2020tnb}, regions previously inaccessible due to computational constraints.
These efforts complement other publicly available numerical relativity
waveform catalogs produced by the community, notably the SXS
catalog~\cite{Boyle:2019kee,Scheel:2025jct}, the Georgia Tech
catalog~\cite{Jani:2016wkt}, and its successor, the MAYA
catalog~\cite{Ferguson:2023vta}, whose recent releases also include
eccentric and precessing configurations.

A critical frontier in current gravitational wave astronomy is the detection and characterization of eccentric binaries. While standard quasi-circular mergers are expected from isolated binary evolution, dynamical formation channels—such as interactions in globular clusters or active galactic nuclei—are predicted to produce binaries with non-negligible eccentricity entering the LIGO frequency band. Accurately modeling these systems requires full numerical relativity simulations that capture the complex modulation of the waveform phase and amplitude.

In this paper, we present the {\it Fifth RIT Binary Black Hole Catalog}. This release adds 248 (plus 25 additional resolutions) new full numerical simulations to our public database, with a specific emphasis on eccentric configurations. With this addition, the RIT Catalog reaches a significant milestone in parameter space coverage: the collection now includes a comparable number of datasets for both orbital regimes, surpassing a thousand quasi-circular and a thousand eccentric simulations. This balanced dataset is designed to support the training of next-generation waveform models and machine learning pipelines capable of distinguishing orbital eccentricity from spin-precession effects, thereby enhancing our ability to infer the formation history of detected sources.


\section{Full Numerical Methods}
\label{sec:methods}

The simulations presented in this catalog were performed using the \texttt{LazEv} code implementation~\cite{Zlochower:2005bj} of the moving puncture formalism~\cite{Campanelli:2005dd, Baker:2005vv}. This section outlines the evolution techniques, the generation of initial data for both quasi-circular and eccentric binaries, and the corrections applied to extrapolate waveforms to infinity and to minimize center-of-mass drift.

\subsection{Evolution of Binary Black Hole Spacetimes}
\label{subsec:evolution}

The \texttt{LazEv} code is built upon the \texttt{Cactus/ETK} computational toolkit~\cite{Schnetter:2003rb} and the \texttt{Carpet} adaptive mesh refinement driver~\cite{Schnetter:2003rb}. We evolve the General Relativity field equations using the BSSNOK formalism~\cite{Nakamura:1987zz, Shibata:1995we, Baumgarte:1998te}, which employs a conformal rescaling of the metric and a trace-free decomposition of the extrinsic curvature to favor numerical stability.

To gauge the spacetime, we utilize the standard moving puncture gauge conditions, consisting of a $1+\log$ slicing condition for the lapse $\alpha$ and a $\Gamma$-driver shift condition for the shift vector $\beta^i$~\cite{Alcubierre:2002kk, Campanelli:2005dd}. The spatial derivatives are computed using eighth-order finite-difference operators~\cite{Lousto:2007rj} to maximize phase accuracy during the long inspiral, while Kreiss-Oliger dissipation is applied to filter high-frequency noise. The time integration is performed using a fourth-order Runge-Kutta method (RK4).

The computational domain consists of a set of nested refinement levels. The finest grids are centered around each black hole, dynamically tracking their motion, while the coarser grids cover the wave extraction zone extending to the outer boundary. We locate the apparent horizons at every time step using the \texttt{AHFinderDirect} thorn~\cite{Thornburg:2003sf}. Gravitational radiation is extracted at various finite radii using the Newman-Penrose scalar $\psi_4$ and subsequently extrapolated to infinity to provide the asymptotic waveforms~\cite{Nakano:2015pta}.

\subsection{Initial Data Generation}
\label{subsec:initialdata}

For the new simulations in this catalog we generated standard initial data using the \texttt{TwoPunctures} spectral solver~\cite{Ansorg:2004ds}, which solves the Hamiltonian and momentum constraints for Bowen-York extrinsic curvature~\cite{Bowen:1980yu}. The input parameters for the solver are the mass parameters, positions, linear momenta, and spins of the two black holes.

\subsection{Quasi-Circular Initial Data (Direct 3.5PN Method)}
\label{subsec:qc_initialdata}

For the majority of the quasi-circular simulations in this catalog, we employ the ``direct'' post-Newtonian (PN) method described in Ref.~\cite{Ciarfella:2024clj}. This approach extends the determination of initial orbital parameters to full 3.5PN accuracy, providing a high-quality ``cold start'' without the computational cost of the iterative reduce-measure-correct loop used in previous works.

The method solves the PN equations of motion to determine the initial tangential momentum $p_t$ and radial momentum $p_r$ required for a quasi-circular inspiral. We utilize the 3.5PN Hamiltonian in the ADM transverse-traceless (ADMTT) gauge, incorporating all spin-orbit, spin-spin, and cubic-in-spin interaction terms. The radial momentum is determined by equating the kinematic radial velocity to the shrinking rate of the orbit due to gravitational wave emission (computed via the 3.5PN energy flux). Similarly, the tangential momentum is derived from the generalized Keplerian relationship for circular orbits in the effective potential.

By including these higher-order corrections, this method directly yields initial data with residual eccentricities of the order $e \sim 10^{-3}$~\cite{Ciarfella:2024clj}. This level of accuracy is sufficient for the bulk production of training data for waveform models, significantly streamlining the simulation pipeline by bypassing the need for pilot evolutions.

\subsection{Eccentric Initial Data Parameters}
\label{subsec:ecc_params}

The initial data for the eccentric simulations in this catalog are generated using the tangential momentum-scaling approach described in Ref.~\cite{Ciarfella:2022hfy}. Rather than performing an iterative search for parameters that yield a specific eccentricity, we systematically explore the parameter space by scaling the tangential linear momentum of a reference quasi-circular configuration. The momenta of the holes is uniquely determined in the Bowen-York initial data set up \cite{Bowen:1980yu}.

For a given binary mass ratio $q$ and initial separation $r$, we first compute the tangential linear momentum required for a quasi-circular orbit, $p_t^{QC}$, using the 3.5PN equations of motion. The eccentric configurations are then defined by introducing a fractional scaling parameter $\text{f} \in (0, 1]$, such that the initial tangential momentum is set to:
\begin{equation}
p_t = (1 - \mathrm{f}) \times p_t^{\rm QC}.
\end{equation}
The radial momentum is typically set to zero ($p_r = 0$), corresponding to an evolution starting at the apastron of the orbit.

To determine the physical eccentricity $e_{3.5\text{PN}}$ associated with a chosen fraction f, we employ the 3.5PN order constitutive equations. By substituting the modified initial momenta and the binary's binding energy into the 3.5PN energy and angular momentum relations, we calculate the effective potential of the orbit. The eccentricity is then derived from the radial turning points (periastron $r_p$ and apastron $r_a$) of this potential via the standard definition:
\begin{equation}
    e_{3.5\text{PN}} = \frac{r_a - r_p}{r_a + r_p}.
\end{equation}
This mapping provides a robust, gauge-invariant link between the initial data control parameter $\mathrm{f}$ (with $\mathrm{f}=0$ for quasicircular orbits and
$\mathrm{f}=1$ for headon) and the physical eccentricity, allowing for the precise categorization of the simulations in the catalog.

Among the advantages of starting the numerical simulations at the apastron one can mention
the fact that in applications to gravitational waves signals, its characterization of the
initial frequency (for instance, at 20Hz) ensures that all the subsequent evolution will
be entirely covered by frequencies $\geq20$Hz, hence in band.
Another advantage, is that since the apastron represents in general a larger initial
separation of the holes and slower speeds, the effects of the 'spurious' radiation
due to the (conformally flat) initial data ansatz are diminished with respect to their
corresponding quasicircular orbits. And, if needed further reduction of this
effect, for instance due to near-maximal spins, we are able to use the HiSpID \cite{Ruchlin:2014zva}
data that ensures at least an order of magnitude reduction of this 'spurious' initial
radiation.


\subsection{Center of Mass Correction (A Posteriori Mode De-Mixing)}
\label{subsec:com_demixing}

While the initial data is corrected to minimize the initial center-of-mass drift, residual linear momentum flux and asymmetric gravitational wave emission during the merger can cause the binary's center of mass to drift away from the coordinate origin. This displacement induces a spurious mixing of the gravitational wave modes $\psi_4^{\ell m}$ measured at finite radii. To restore the fidelity of the waveform, we apply an \textit{a posteriori} correction described in Section II.A of Ref.~\cite{Healy:2020vre}, often referred to as ``mode de-mixing.''

The correction assumes that the extracted waveform $\psi_4(t, \vec{r})$ corresponds to a source that is translated by a time-dependent vector $\vec{\xi}(t)$ relative to the origin. This translation mixes the pure spin-weighted spherical harmonic modes. To recover the modes in the source's center-of-mass frame, we apply the inverse transformation. The corrected modes $\psi_4^{\prime \ell m}$ are constructed from the measured modes $\psi_4^{\ell m}$ via a mixing formula:
\begin{equation}
    \psi_4^{\prime \ell m} = \sum_{\ell', m'} \mathcal{C}_{\ell m}^{\ell' m'}(\vec{\xi}) \psi_4^{\ell' m'},
\end{equation}
where the coefficients $\mathcal{C}$ depend on the displacement vector $\vec{\xi}(t)$.

The trajectory of the center of mass $\vec{\xi}(t)$ is determined directly from the waveform itself. By minimizing the power in the non-radiative dipole modes (or physically motivated ``unphysical'' mode content) or by integrating the momentum flux conservation equations, we isolate the drift. This post-processing step effectively ``de-mixes'' the signal, ensuring that the final catalog waveforms represent the radiation field as emitted from the center of mass, thereby maximizing the accuracy of higher-order mode amplitudes for data analysis.


\subsection{Waveform Extraction and Extrapolation}
\label{subsec:extrapolation}

Gravitational waves are extracted from the numerical grid by computing the Weyl scalar $\psi_4$ on a series of concentric spheres at finite radii $r_{obs}$ from the source. While these measurements provide a direct readout of the radiation field, they are contaminated by near-field effects ($ \sim O(1/r)$) and gauge ambiguities that vanish only asymptotically. To obtain high-precision waveforms suitable for gravitational wave data analysis, it is necessary to extrapolate these signals to null infinity ($\mathscr{I}^+$).

In this catalog, we employ the perturbative extraction technique derived in Ref.~\cite{Nakano:2015pta}. Rather than relying solely on polynomial extrapolation of the raw data in powers of $1/r$, this method utilizes an analytical expression based on the Teukolsky equation to propagate the waveform from a finite radius to infinity. By modeling the propagation of the wave through the curved background of the final remnant black hole, we apply corrections that account for the peeling theorem and the specific fall-off behavior of the Weyl scalars.

The corrected waveform at infinity, $\psi_4(t,r\to\infty)$, is constructed from the finite-radius measurement $\psi_4(t, r_{obs})$ via a series expansion that filters out the leading-order $1/r$ potentials and gauge effects. This approach has been shown to significantly reduce phase and amplitude errors compared to standard polynomial extrapolation, particularly for observers located at moderate distances ($r_{obs} \sim 100M-200M$) where the computational grid is most resolved. We typically use $r_{obs}=113M$ for quasicircular orbits and $r_{obs}=240M$ for highly eccentric ones \cite{Ficarra:2024nro}.

\section{New studies of eccentric merging binaries}\label{sec:eccentricity}

In this new release of the RIT catalog, while we keep the simplicity of its format, we have incorporated a few representative cases of waveform convergence for long, eccentric binaries simulations, since all our previous convergence studies rely on quasicircular orbits.
While the numbering NNNN of the catalog cases, RIT:BBH:NNNN, is based on the initial intrinsic parameters of the binary (note that from 
this fifth release on we have dropped the 'e' from the former
RIT:eBBH: notation), we additionally identify them with the nxxx resolution label at the extraction radii.
We have also incorporated a much more accurate estimate of the initial eccentricity, i.e. $e$3.5PN over the previous release that used the Newtonian estimate $e_0$
(see some comparative examples in Table VI of \cite{McMillin:2025hof}). All waveforms released include up to the modes $\ell=4$, and strain up to
$\ell=5$ for eccentric cases, (even if we extract up to $\ell=6$ in current simulations).

\begin{figure*}[ht!]
    \includegraphics[angle=0,width=1.11\columnwidth]{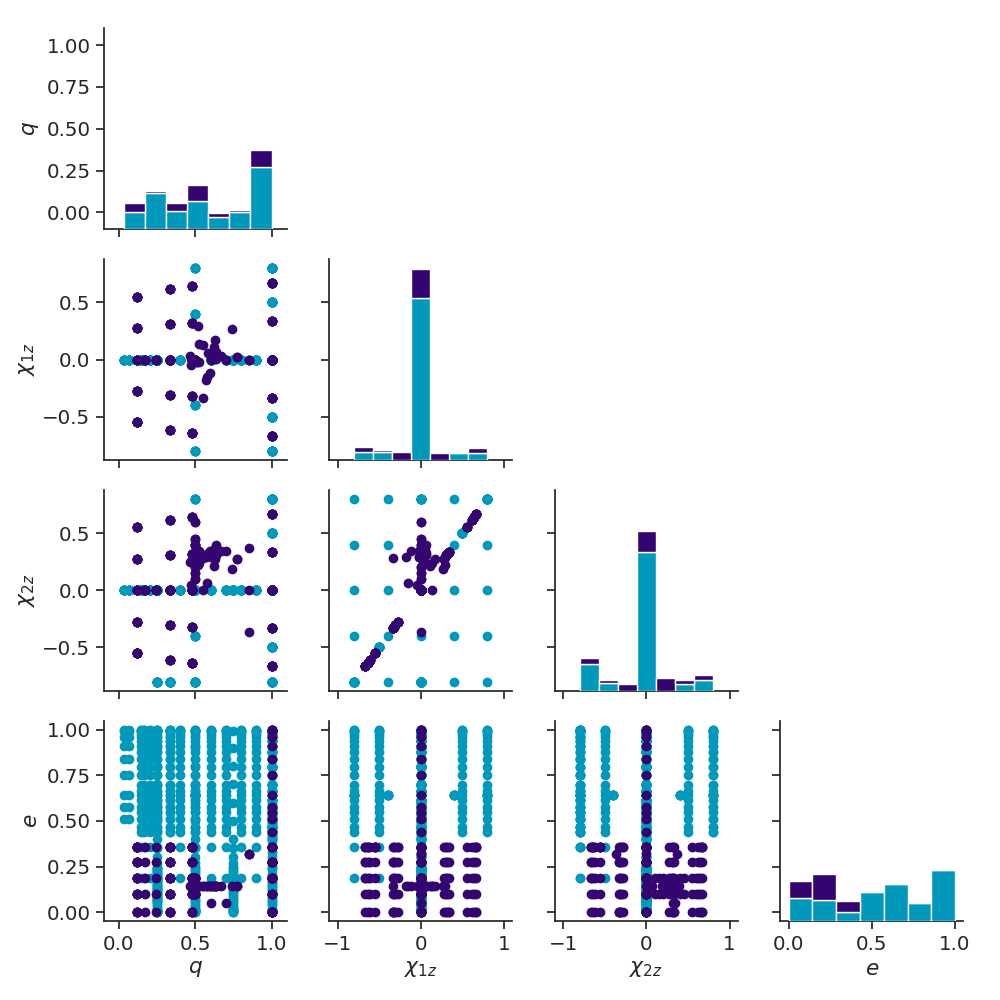}
  \caption{Counting simulations in the $(q,\chi_1^z,\chi_2^z,e)$ planes 
    (faces of the 4-cube) for the 906 nonprecessing eccentric binaries.
    The 709 of release 4 simulations are in red, and the 197 of release 5 in blue.
      \label{fig:AlignedMulti4}}
\end{figure*}

As in the fourth release we started to include eccentric simulations based on our motivation to estimate parameters for the GW190521 event \cite{Gayathri:2020coq}, in this new release the new eccentric simulations have been in part motivated by the study of the events GW200208\_22 and GW190620 \cite{McMillin:2025hof}. As a total result, in this fifth release we have reached a sort of balance between the eccentric and noneccentric simulations reaching each over one thousand cases.
We note, however, that since eccentricity adds an extra dimension to the
binary parameter space, an equal number of simulations does not imply an
equal coverage density of the eccentric and quasicircular sectors; rather,
it represents a first step towards reducing the sampling bias in favor of
the noneccentric simulations when applying our catalog to the parameter
estimation of GW events. 
Table~\ref{tab:Catalogs} gives a summary of the five releases contributions and Figure~\ref{fig:AlignedMulti4} shows an overview of the relative contributions to aligned eccentric binaries parameters from releases 4 and 5, and the now more homogeneous coverage of the aligned parameter space.

\begin{table*}
\caption{Summary of simulations in each release of the RIT catalog}
\label{tab:Catalogs}
\begin{ruledtabular}
  \begin{tabular}{|l|c|c|c|c|c||c|}
Release/[cite] & \#1\cite{Healy:2017psd} & \#2\cite{Healy:2019jyf} & \#3\cite{Healy:2020vre} & \#4\cite{Healy:2022wdn} & \#5\cite{Ficarra:2026cej}  & Total\\
\hline
QC Nonspinning & 21 & 4 & 4 & 14 & 1 & 44 \\
QC Spinning Aligned  & 99 & 150 & 199 & 120 & 13 & 581 \\
QC Precessing  & 6 & 40 & 254 & 146 & 37 & 483 \\
Eccentric Nonspinning & - & - & - & 511 & 65 & 576 \\
Eccentric Spinning Aligned  & - & - & - & 198 & 132 & 330 \\
Eccentric Precessing  & - & - & - & 115 & - &115 \\
\hline
Subtotal   & 126 & 194 & 457 & 1104 & 248 & 2129 \\
\end{tabular}
\end{ruledtabular}
\end{table*}

\subsection{Convergence of the Numerical Simulations}
\label{subsec:convergence}

To assess the accuracy and reliability of our eccentric binaries numerical evolutions, we conducted a systematic convergence study of
chosen configurations using three distinct grid resolutions, systematically lowering or raising our standard grid
resolution by successive factors of $f=1.2$, where
$f=h_L/h_M=h_M/h_H$ denotes the
ratio of grid spacings between successive global resolutions.
In our notation, resolutions are labeled as ``nXXX'',
where the numerical value indicates the resolution within the wave extraction zone.
For instance, the standard baseline resolution used for the rest of the simulations is denoted as n120 (corresponding to a spatial resolution of $M/1.20$ in the wavezone), while the coarser resolution used in the convergence test is n100 ($M/1.00$).

Assuming that any extracted waveform quantity $\Psi$ depends on the grid spacing $h$ following the relation $\Psi(h) = \Psi_\infty + A h^n$, we compared the results across the low ($h_L$), medium ($h_M$), and high ($h_H$) resolution runs.
Finally, by utilizing the data from the three resolutions, we applied Richardson extrapolation to estimate the exact theoretical limit at infinite resolution, $\Psi_{\infty}$ (for either the $\psi_4$ or the strain $h$ itself), as given in
Eqs.~(5a-5c) of \cite{Lousto:2019lyf}.

\begin{subequations}
\begin{align}
\Psi_{\infty} &= \frac{\Psi_{H}\Psi_{L} - \Psi_{M}^{2}}{\Psi_{H} - 2\Psi_{M} + \Psi_{L}}, \label{eq:5a} \\
<A>&= \frac{\Psi_{L}^{2} - 2\Psi_{L}\Psi_{M} + \Psi_{M}^{2}}{\Psi_{H} - 2\Psi_{M} + \Psi_{L}}, \label{eq:5b} \\
n &= -\frac{1}{\ln(f)} \ln \left| \frac{\Psi_{H} - \Psi_{M}}{\Psi_{L} - \Psi_{M}} \right|. \label{eq:5c}
\end{align}
\end{subequations}

We have applied these estimates to binary black holes in quasicircular orbits in previous works and now
we give in the appendix \ref{sec:convergence} here the results of applying them to the eleven
selected eccentric binaries configurations.

From these relationships we can also derive an estimate of the relative errors explicitly as
\begin{subequations}
\begin{align}
    \Delta_\infty^H&=\frac{(\Psi_\infty-\Psi_H)}{\Psi_\infty}=\frac{(\Psi_H-\Psi_M)^2}{(\Psi_M^2-\Psi_H\Psi_L)},\\
    \Delta_\infty^M&=\frac{(\Psi_\infty-\Psi_M)}{\Psi_\infty}=\frac{(\Psi_H-\Psi_M)(\Psi_M-\Psi_L)}{(\Psi_M^2-\Psi_H\Psi_L)},\\
    \Delta_\infty^L&=\frac{(\Psi_\infty-\Psi_L)}{\Psi_\infty}=\frac{(\Psi_M-\Psi_L)^2}{(\Psi_M^2-\Psi_H\Psi_L)}.
\end{align}
\end{subequations}


\subsection{Waveform error estimates}\label{sec:waveforms}

\begin{figure*}[ht!]
    \centering
    \includegraphics[width=.75\textwidth]{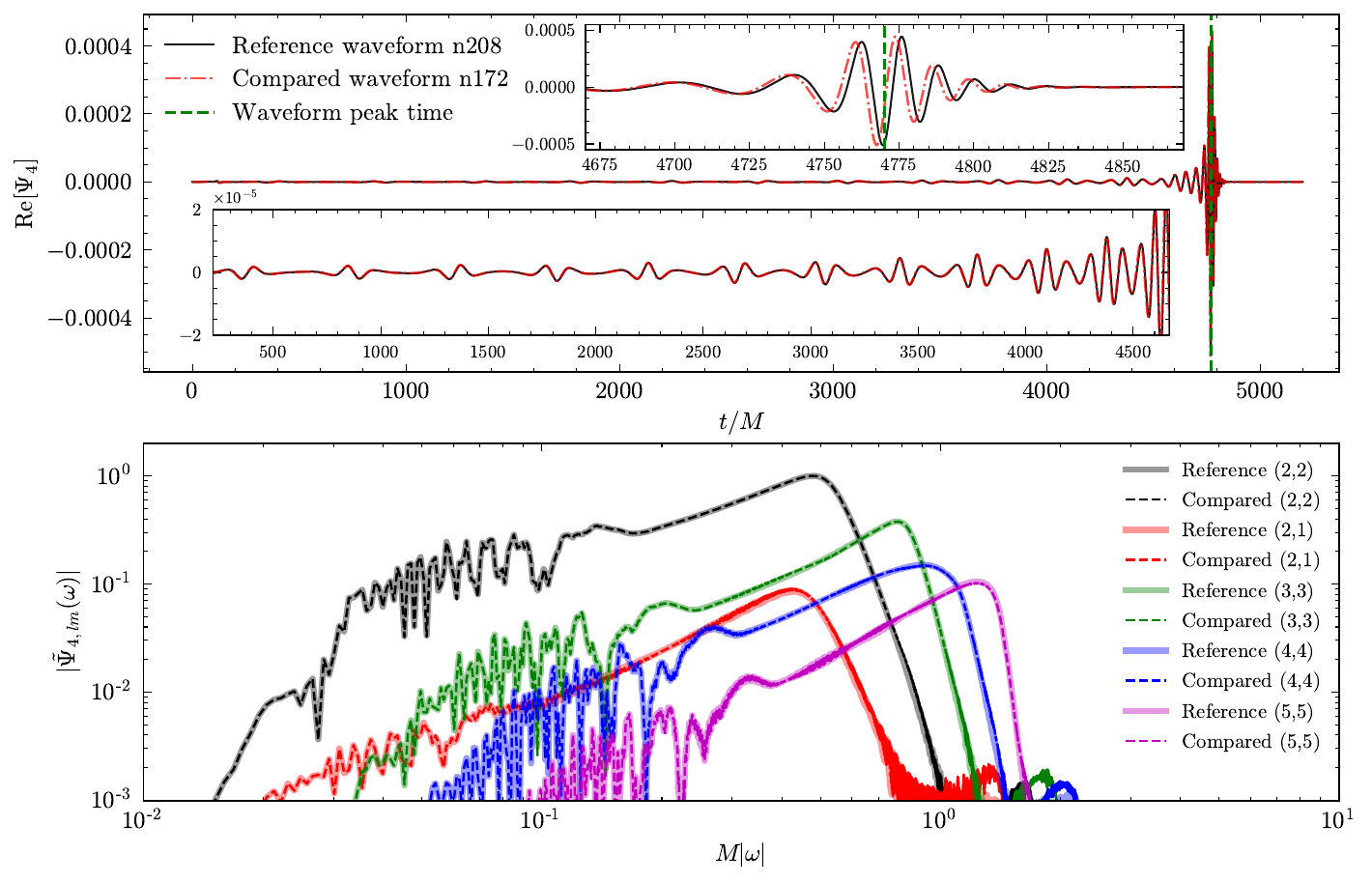}
    \caption{Representative example used for the computation of the
      mismatch in Tab.\ref{tab:convergence_mismatch_rit_catalog5_sixres_runs} between the n172 and n208 cases,
      including all modes up to $(\ell,m)=(4,4)$ and integrating over both positive and negative
      frequencies with no frequency cutoff (here ${\cal M}(\Psi_{n208},\Psi_{n172})=1.09\times10^{-4}$).
      The lower panel shows the amplitude spectra $|\tilde\psi_4^{\ell m}(\omega)|$ of the individual modes for
      both resolutions.} 
    \label{fig:matching_example}
\end{figure*}

We provide here a direct comparison of the waveforms 
at different resolutions by computing their overlap over the whole range of the 
simulations as given by the mismatching measure,
\begin{eqnarray}
          {\cal M}\equiv 1-\frac{\left<h_1\left|\right.h_2\right>}{\sqrt{\left<h_1\left|\right.h_1\right>\left<h_2\left|\right.h_2\right>}},\label{eq:mismatch}
\end{eqnarray}
as implemented via a complex overlap as described in Eq.~(2) in 
Ref.~\cite{Cho:2012ed}:
\begin{equation}
  \left< h_1 \left|\right. h_2 \right>=4\,\Re\int_{-\infty}^{\infty} \frac{d\omega}
         {S_n(\omega)}\left[\tilde{h}_1(\omega) \tilde{h}_2(\omega)^* \right],\label{eq:match}
\end{equation} 
where $\tilde{h}(\omega)$ is the Fourier transform of $h(t)$ and $S_n(\omega)$ 
is the power spectral density of the detector noise (here specifically taken 
as white noise $S_n(\omega)=1$ since we are interested in the direct waveform
comparisons, the effective faithfulness thresholds for actual LVK parameter
estimation will be weighted by the detectors' frequency-dependent sensitivities).
We include in the computation all the modes up to $(\ell,m)=(4,4)$ of $\psi_4$,
adopting the mode-summed mismatch in which both the inner product and the norms
of Eq.~(\ref{eq:mismatch}) are summed over the modes,
\[
{\cal M} = 1 - \max_{\Delta t,\,\Delta\phi}
\frac{\sum_{\ell m}\left< h_1^{\ell m}\left|\right. h_2^{\ell m}\,e^{i m \Delta\phi} \right>}
{\sqrt{\sum_{\ell m}\left< h_1^{\ell m}\left|\right. h_1^{\ell m}\right>\;
       \sum_{\ell m}\left< h_2^{\ell m}\left|\right. h_2^{\ell m}\right>}},
\]
maximized over a common time shift $\Delta t$ and rigid orbital-phase rotation
$\Delta\phi$ (under which each mode transforms as $h^{\ell m}\to h^{\ell m}e^{im\Delta\phi}$),
so that each mode enters weighted by its power relative to the
leading $(2,2)$ mode. By the orthonormality of the spin-weighted spherical
harmonics, this definition is equivalent to the mismatch of the observer-frame
waveform, $h(t;\iota,\phi_0)=\sum_{\ell m}\,{}_{-2}Y_{\ell m}(\iota,\phi_0)\,h^{\ell m}(t)$,
averaged over all source orientations; the modes are thus effectively recombined
with their spherical-harmonic weights over the whole sphere, accounting in an
averaged sense for the increased relative contribution of the subdominant modes
at large inclinations. This coincides with the convention adopted
for numerical relativity error estimates in the SXS catalog and in the
construction of numerical relativity surrogate models
\cite{Boyle:2019kee,Blackman:2017pcm,Varma:2019csw}.
Mismatches are computed using our python scripts, that maximizes the match
integral over constant time and phase shifts directly in Fourier space
(See also \cite{Allen:2005fk}).
Before doing that, we apply a Planck taper at the start and at the end of
the complete signals in the time-domain \cite{McKechan:2010kp},
in order to make it periodic so
it does not spoil the discrete Fourier transform.
Since the individual $\psi_4^{\ell m}$ modes are complex, we integrate over both
positive and negative frequencies, $-\infty<\omega<\infty$, without imposing any
frequency cutoff, and we maximize over an overall constant time shift and an
overall constant phase shift. An example of this procedure is displayed in
Fig.~\ref{fig:matching_example}

Using the convergence formulas Eqs.~(\ref{eq:5a})-(\ref{eq:5c}) above, we can compute the
mismatches of each of the waveforms with respect to that extrapolated to infinity
(without needing to compute it explicitly) and provide a measure of their expected accuracy.
Since the mismatch is quadratic in the waveform difference,
${\cal M}(\Psi_X,\Psi_Y)=\tfrac12\|\hat\Psi_X-\hat\Psi_Y\|^2$ for unit-normalized, aligned waveforms,
the Richardson relations hold for $\sqrt{\cal M}$, leading to
\begin{equation}\label{eq:MH}
{\cal M}(\Psi_\infty,\Psi_H)=\frac{{\cal M}(\Psi_H,\Psi_M)^2}{
\left[\sqrt{{\cal M}(\Psi_L,\Psi_M)}-\sqrt{{\cal M}(\Psi_H,\Psi_M)}\right]^2},
\end{equation}
and also
\begin{equation}\label{eq:MM}
{\cal M}(\Psi_\infty,\Psi_M)=\frac{{\cal M}(\Psi_H,\Psi_M)\,{\cal M}(\Psi_L,\Psi_M)}{
\left[\sqrt{{\cal M}(\Psi_L,\Psi_M)}-\sqrt{{\cal M}(\Psi_H,\Psi_M)}\right]^2},
\end{equation}
and
\begin{equation}\label{eq:ML}
{\cal M}(\Psi_\infty,\Psi_L)=\frac{{\cal M}(\Psi_L,\Psi_M)^2}{
\left[\sqrt{{\cal M}(\Psi_L,\Psi_M)}-\sqrt{{\cal M}(\Psi_H,\Psi_M)}\right]^2}.
\end{equation}
where we used that
\begin{equation}\label{eq:fn}
    f^n={\cal M}(\Psi_L,\Psi_M)/{\cal M}(\Psi_H,\Psi_M),
\end{equation}
and hence the convergence order can be obtained as,
\begin{equation}\label{eq:n}
    n=\ln\left|{\cal M}(\Psi_L,\Psi_M)/{\cal M}(\Psi_H,\Psi_M)\right|/\ln(f).
\end{equation}
Note that $n$ so defined is the convergence order of the mismatch, which is twice that
of the waveform itself. These relations assume that the resolution error is dominated by
a single term, $\Psi_X-\Psi_\infty\simeq A\,h_X^{n/2}\,\varepsilon(t)$ with a common
$\varepsilon(t)$, so that the error vectors of the three resolutions are parallel and
${\cal M}(\Psi_X,\Psi_Y)\propto(h_X^{n/2}-h_Y^{n/2})^2$; a direct check of this assumption,
together with a more complete error analysis, will be presented elsewhere.

The ``faithfulness'' or ``indistinguishability'' SNR ($\rho_{\rm faith}$)
at which a given mismatch $\mathcal{M}$ between a true signal $s$ and a
template $h$ (or two different models $h_1,h_2$) becomes problematic or
results in parameter bias is approximately related by the following
formula (based on the linear signal approximation):
$\rho_{\rm faith}\approx\frac{\sqrt{k_{df}/2}}{\sqrt{\mathcal{M}}}$
\cite{Lindblom:2008cm}, where $k_{df}$ is the quantile of the $\chi^2$
distribution with number of degrees of freedom equal to the number of
parameters being measured. For a single-parameter measurement at the
90\% credible level, $k_{df}\approx2.71$, leading to the effective
conservative faithfulness criterion
$\rho\approx\sqrt{1.35}/\sqrt{\mathcal{M}}$ \cite{Mahapatra:2026wsp}
that we report in practice.

For the sake of completeness, we mention here that in the case of two waveforms the mismatch takes the form
\begin{equation}
{\cal M}^H(\Psi_\infty,\Psi_H)={\cal M}(\Psi_H,\Psi_M)/(f^{n/2}-1)^2,
\end{equation}
and
\begin{equation}
{\cal M}^M(\Psi_\infty,\Psi_M)=f^n{\cal M}^H(\Psi_\infty,\Psi_H).
\end{equation}
Notably, for $f=1.2$ the factor $1/(f^{n/2}-1)^2<1$ for $n>7.6$, thus simply 
${\cal M}(\Psi_H,\Psi_M)$ provides a bound to the mismatch of the high-resolution run
with respect to the extrapolation to infinity if the convergence order is $n>7.6$.
This is a sufficient condition for using the pair mismatch as an upper bound without
extrapolation, not a condition for convergence: for $0<n<7.6$ the sequence still converges,
but ${\cal M}(\Psi_H,\Psi_M)$ underestimates the error of the high-resolution run by the
factor $1/(f^{n/2}-1)^2$ (e.g.\ $3.4$ for $n=6.6$), which is why the extrapolation of
Eqs.~(\ref{eq:MH})-(\ref{eq:ML}) is used.
Note that upon introducing expression (\ref{eq:fn}) into the above equations we
recover expressions (\ref{eq:MH})-(\ref{eq:MM}).

The mismatches reported in
Tables~\ref{tab:convergence_mismatch_rit_catalog5_giuseppe_runs} and
\ref{tab:convergence_mismatch_rit_catalog5_sixres_runs} include all the modes up to
$(\ell,m)=(4,4)$, coherently summed as described above, so that the quoted values and
faithfulness SNRs characterize the accuracy of the full higher-mode waveform rather
than that of the dominant $(2,2)$ mode alone. We find that including the higher modes
changes the mismatches by at most a factor of a few relative to a $(2,2)$-only
computation, and leaves the overall convergence picture---and in particular the
conclusion that the standard medium resolution (n120) lies in a convergent regime with
bounded errors for the bulk of the catalog---unchanged. The largest relative degradation occurs, as
expected, for the most demanding configurations (small mass ratios and large
antialigned spins), such as RIT:BBH:2104 and RIT:BBH:2146, where the higher-mode
content is comparatively more significant.

In Table~\ref{tab:convergence_mismatch_rit_catalog5_giuseppe_runs} we report the results
of applying this method to estimate the accuracy of the low (L), medium (M), and
high (H) resolution
waveforms produced by the merger of eccentric merging binary black holes.
We compute the effective
convergence power $n$ directly from expression (\ref{eq:n}) above. This serves to
control the convergence of the procedure. We also provide as a measure of the reference
minimal signal-to-noise-ratio (SNR) these waveforms are considered accurate for parameter
estimation,
$\rho_{L,M,H}\sim\sqrt{1.35}/\sqrt{\cal M}$, for the low, medium, and high resolutions.
We have added to the description of the simulations their estimated number
of orbits to merger to provide a length reference. More details of the individual
simulations are left to the appendices here.

The top
four set of simulations reported in this table actually correspond to configurations
already reported in the release 4 of our catalog. We revisit them here with the addition
of two different resolutions to complete the three sets of n100, n120, n144 corresponding
respectively to low, medium, and high resolutions.
While two of the simulations RIT:BBH:1637 and RIT:BBH:1670 essentially represent
plunging binaries the other two RIT:BBH:1282 and RIT:BBH:1283 provide lengthy
eccentric evolutions. We note that the now standard medium resolution (n120) simulations
of these four configurations, as well as of the release 5 cases RIT:BBH:2013 and the
RIT:BBH:2010 sequence discussed below, yield $\rho_M\approx20$--$60$, in line with the SNR
of typical LVK signals. For the more demanding configurations RIT:BBH:2146, RIT:BBH:2170,
and RIT:BBH:2172 the medium resolution gives $\rho_M\lesssim8$ and higher resolutions are
required, while for RIT:BBH:2104 and RIT:BBH:2153 the three resolutions are not yet in the
convergent regime ($n<3$) and no extrapolation is meaningful. The low $\rho_{L}$ values and
the overall high power $n$ of RIT:BBH:2013, RIT:BBH:2146, and RIT:BBH:2172 are indicative
that their n100 simulations are just entering the convergence regime.

\begin{table*}[ht!]
  \caption{Convergence of mismatch for several convergence runs from release 5 of the RIT Catalog.
    We use resolutions $n100$, $n120$ and $n144$ and a factor $f=1.2$. The mismatches include all modes up to $(\ell,m)=(4,4)$ and are integrated over both positive and negative frequencies with no frequency cutoff. We also report the values of the effective faithfulness $\rho \approx \color{black}\sqrt{1.35}/\sqrt{\cal M}$ \cite{Mahapatra:2026wsp}.
$^{a}$The mismatch between the n100 and n120 runs themselves (not tabulated) is ${\cal M}(\Psi_L,\Psi_M)=0.38$, $0.83$ and $0.94$ for RIT:BBH:2013, 2146 and 2172, respectively: the low-resolution run lies outside the regime of validity of the small-mismatch expansion, the extrapolated values in these rows are formal, and entries exceeding unity are omitted.
$^{b}$$n<3$, i.e.\ ${\cal M}(\Psi_L,\Psi_M)\approx{\cal M}(\Psi_H,\Psi_M)$: the three resolutions are not yet in the convergent regime and no extrapolation is meaningful.
\label{tab:convergence_mismatch_rit_catalog5_giuseppe_runs}
}
\centering
\begin{tabular}{lcccccccc}
\hline
Tag & $N_{orbits}$ & $n$ & ${\cal M}(\Psi_\infty,\Psi_L)$ & ${\cal M}(\Psi_\infty,\Psi_M)$ & ${\cal M}(\Psi_\infty,\Psi_H)$ & $\rho_L$ & $\rho_{M}$ & $\rho_H$ \\
\hline
RIT:BBH:1282 &31.58& 6.07 & 2.5179e-03 & 8.3191e-04 & 2.7487e-04 & 23.16 & 40.28 & 70.08 \\
RIT:BBH:1283 &20.94& 5.00 & 2.8311e-03 & 1.1374e-03 & 4.5692e-04 & 21.84 & 34.45 & 54.36 \\
RIT:BBH:1637 &1.04& 6.04 & 3.1749e-03 & 1.0544e-03 & 3.5017e-04 & 20.62 & 35.78 & 62.09 \\
RIT:BBH:1670 &1.23& 6.48 & 2.0336e-03 & 6.2318e-04 & 1.9097e-04 & 25.77 & 46.54 & 84.08 \\
\hline
RIT:BBH:2013$^{a}$ &17.72& 38.82 & 4.0648e-01 & 3.4315e-04 & 2.8969e-07 & 1.82 & 62.72 & 2158.74 \\
RIT:BBH:2104$^{b}$ &22.46& 0.53 & 4.2229e-01 & 3.8331e-01 & 3.4793e-01 & 1.79 & 1.88 & 1.97 \\
RIT:BBH:2146$^{a}$ &12.51& 10.14 & --- & 3.6024e-01 & 5.6702e-02 & --- & 1.94 & 4.88 \\
RIT:BBH:2153$^{b}$ &21.73& 1.49 & 1.6239e-01 & 1.2361e-01 & 9.4084e-02 & 2.88 & 3.30 & 3.79 \\
RIT:BBH:2170 &20.05& 6.79 & 7.4600e-02 & 2.1642e-02 & 6.2787e-03 & 4.25 & 7.90 & 14.66 \\
RIT:BBH:2172$^{a}$ &25.82& 15.06 & --- & 1.0769e-01 & 6.9108e-03 & --- & 3.54 & 13.98 \\
\hline
\end{tabular}
\end{table*}

In Table~\ref{tab:convergence_mismatch_rit_catalog5_sixres_runs} we report the results
of a six resolutions study of RIT:BBH:2010, an eccentric $(e\sim0.25)$ simulation
with 18 orbits to merger. This configuration was of particular interest
since it represented the top match to the LVK event GW200208\_22 as studied in
\cite{McMillin:2025hof}. In this paper we have seen that different finite differences
resolutions had very little effect on the matching likelihood $\lnL$ (see table III
of \cite{McMillin:2025hof}). Here we take the opportunity to study the convergence
properties and estimated errors by including a very low global resolution (n084)
and higher resolutions than usual, n172 and n208 corresponding (approximately) to
further increases by factors of $f\sim1.2$ to study the approach to the convergence
regime in successive sequences of three resolutions, as displayed in the top
four combinations. We observe convergence orders $n\approx6$--$7$ of the mismatch,
corresponding to waveform convergence orders of $3$--$3.7$, between the second-order
prolongation and the fourth-order time integration, except for the n120-n144-n172
combination that produces three close to each other values of the mismatches.

These error fluctuations can be controlled and reduced
by taking larger differences between the global resolutions, for instance, skipping
every other as in n100, n144, n208 gives an effective $f=1.2^2=1.44$ that leads to
a cleaner convergence and SNR in line with expected values and with satisfying LVK standards.
We have also done this for the n084, n120, n172 combination for a further control of the
methods, obtaining very satisfactory convergence rates and values of their corresponding
$\rho_{L,M,H}\sim\sqrt{1.35}/\sqrt{\cal M}$.

\begin{table*}[ht!]
  \caption{Convergence of mismatch for RIT:BBH:2010 from release 5 of the RIT Catalog using all available resolutions with an $N_{orbits}\sim18$.
    We consider a factor $f=1.2$ for all combinations except for the last two rows where we use $f=1.44$; these values are nominal, the actual successive resolution ratios differing from them by less than 1\%. As in Table~\ref{tab:convergence_mismatch_rit_catalog5_giuseppe_runs}, the mismatches include all modes up to $(\ell,m)=(4,4)$ and are integrated over both positive and negative frequencies with no frequency cutoff. We also report the values of the effective faithfulness $\rho \sim \color{black}\sqrt{1.35}/\sqrt{\cal M}$ \cite{Mahapatra:2026wsp}. Footnote $^{b}$ as in Table~\ref{tab:convergence_mismatch_rit_catalog5_giuseppe_runs}.
\label{tab:convergence_mismatch_rit_catalog5_sixres_runs}
}
\centering
\begin{tabular}{lccccccc}
\hline
Combination & $n$ & ${\cal M}(\Psi_\infty,\Psi_L)$ & ${\cal M}(\Psi_\infty,\Psi_M)$ & ${\cal M}(\Psi_\infty,\Psi_H)$ & $\rho_L$ & $\rho_{M}$ & $\rho_H$ \\
\hline
084-100-120 & 6.32 & 3.7398e-02 & 1.1813e-02 & 3.7312e-03 & 6.01 & 10.69 & 19.02 \\
100-120-144 & 6.58 & 1.1141e-02 & 3.3581e-03 & 1.0122e-03 & 11.01 & 20.05 & 36.52 \\
120-144-172$^{b}$ & 2.72 & 1.4167e-02 & 8.6291e-03 & 5.2558e-03 & 9.76 & 12.51 & 16.03 \\
144-172-208 & 7.35 & 1.7460e-03 & 4.5748e-04 & 1.1987e-04 & 27.81 & 54.32 & 106.12 \\
\hline
100-144-208 & 3.37 & 6.2719e-03 & 1.8314e-03 & 5.3477e-04 & 14.67 & 27.15 & 50.24 \\
084-120-172 & 9.21 & 1.4651e-02 & 5.0814e-04 & 1.7624e-05 & 9.60 & 51.54 & 276.77 \\
\hline 
\end{tabular}
\end{table*}


\section{Conclusions and Discussion}\label{sec:Discussion}

In this fifth release, we have included a series of longer-term eccentric BBH evolutions utilizing improved estimates for their initial eccentricity.
These additions provide better coverage of the source parameter space necessary for matching LVK gravitational-wave events—particularly for lower total-mass systems,
which allow more of their eccentric orbital features to be observed in-band. Direct application of the expanded catalog to the parameter estimation of GW events, specifically the search for eccentric binary candidates, is currently underway \cite{McMillin:2026}.
Our convergence analysis, based on the computation of waveform mismatches, indicates that the standard medium resolution (n120) is in a convergent regime and yields bounded errors for the bulk of the catalog.
Furthermore, the a posteriori center-of-mass drift corrections applied to this catalog ensure high fidelity not only for the dominant (2,2) mode but also for the higher-order modes. Nevertheless, targeted higher resolutions (n144 and above) will be required for highly demanding binary configurations (such as small mass ratios or high spins), or to meet the stringent accuracy requirements of new-generation detectors.
In future releases, we plan to incorporate scattering events and triple systems. While we have also simulated numerous high-energy collisions \cite{Healy:2022jbh,Healy:2024lhl}, we have opted to exclude them from the current RIT public catalog to maintain its primary focus on parameter estimation for standard astrophysical GW events.

\textit{Note added.}---This version incorporates the corrections of the Erratum to the
published article [Phys.\ Rev.\ D \textbf{114}, 064095 (2026)]: Eqs.~(\ref{eq:MH})-(\ref{eq:ML})
and the two-waveform relation following them, the values of the extrapolated mismatches and
faithfulness SNRs in Tables~\ref{tab:convergence_mismatch_rit_catalog5_giuseppe_runs}
and \ref{tab:convergence_mismatch_rit_catalog5_sixres_runs}, and the accompanying discussion
in Sec.~\ref{sec:waveforms}. The values of $n$ in those tables, and all other results of the
paper, are unchanged.

\begin{acknowledgments}
GF gratefully acknowledges the support of University of Calabria through a research fellowship funded by DR 1688/2023.
COL gratefully acknowledges support from NSF awards AST-2319326, PHY-2207920 and PHY-2513442.
Computational resources were also provided by the BlueSky
Clusters, Green Prairies, and White Lagoon at the Rochester Institute
of Technology, which were supported by NSF grants 
No.\ PHY-1229173, No.\ PHY-1726215, and No.\ PHY-2018420. 
This work also used the ACCESS computational resources from the
allocation PHY060027 and Frontera at TACC project PHY20007.
\end{acknowledgments}

\newpage

\bibliographystyle{apsrev4-1}
\bibliography{references}


\appendix



\section{Convergence studies}\label{sec:convergence}

We have selected ten of the eccentric simulations to perform the convergence studies.
These studies present the convergence of number of orbits $N$ (as measured by the winding of
the punctures trajectories in coordinate space), merger time $t_{\rm{m}}/M$ (as given by coordinate
separation $r/M=0.7$, corresponding closely to the formation of a common horizon and
the peak amplitude of the waveform), remnant mass $M_f$, spin $\chi_f$
(as measured by the final hole isolated horizon area and Killing field
respectively \cite{Dreyer02a,Campanelli:2006fy} ), and recoil $V_f$, 
peak luminosity $\mathcal{L}_{\rm{peak}}$ (erg/s for $50M_\odot$), as directly computed from the
waveforms (adding up to $\ell=6$ modes). While the 
peak frequency $M\omega_{22}^{\rm{peak}}$, and
strain peak amplitude $\left(r/M\right)|h_{22}^{\rm{peak}}|$ are computed from the leading $(\ell,m)=(2,2)$
model peak values.

Richardson extrapolation is used to determine convergence order $n$
and extrapolated values at $n_\infty$.
While our theoretical expectations are convergence orders to lie between the fourth order Runge-Kutta
time integration (though there is a second order prolongation effect)
and the eight order spatial finite difference, the leading error order modeling $(Ah^n)$ can have some deviations
due to particular oscillations in the coefficient $A$ not being constant or higher powers order corrections
$h^{n+i}$ could be added to the modeling. 
Given the increases in global resolution factors being just 1.2 we do not expect that quantities that
agree on the first 3-4 significant digits (within $<0.1\%$) to provide an accurate extrapolation order.
When agreement to 3-4 the convergence order would formally be infinite and we give $n=  - - $ in tables
as well as when $n<1$. Nevertheless in most 
cases the results remain consistent to 6-digits and are given in the tables.

From the catalog's fourth release simulations we have considered
RIT:BBH:1282, RIT:BBH:1283, RIT:BBH:1637, RIT:BBH:1670, covering
from equal mass nonspinning displayed in Tables~\ref{tab:convergence_Jim1}
and~\ref{tab:convergence_Jim2} to highly spinning $\chi_i=0.7$ precessing
in a near plunge configuration as displayed in
Tables~\ref{tab:convergence_Jim3} and~\ref{tab:convergence_Jim4}.

\begin{table*}
\caption{Convergence of number of orbits $N$, merger time $t_{\rm{m}}/M$, remnant mass $M_f$, spin $\chi_f$ and recoil $V_f$, 
peak luminosity $\mathcal{L}_{\rm{peak}}$,  peak frequency $M\omega_{22}^{\rm{peak}}$, and
strain peak amplitude $\left(r/M\right)|h_{22}^{\rm{peak}}|$ for the RIT:BBH:1282 configuration with initial separation $D=24.64 M$, $\chi_{1,2} = 0.0$, $q=1.0$ and $\mathrm{f}=0.10$. Richardson extrapolation is used to determine convergence order $n$ and extrapolated values at $n_\infty$. 
}
\label{tab:convergence_Jim1}

\centering
\begin{tabular}{cccccccccc}
\hline
resolution  & $t_{\rm{m}}/M$& $N$ & $M_f/M$ & $\chi_f$ & $V_f\,$[km/s] & $\mathcal{L}_{\rm{peak}}\,$[erg/s] 
  &$M\omega_{22}^{\rm{peak}}$ &$\left(r/M\right)|h_{22}^{\rm{peak}}|$ \\
\hline
n100 & 11764.3 & 31.68 & 0.9513 & 0.6862 & 0.00 & 3.698e+56 & 0.3571 & 0.3951 \\
n120 & 11733.6 & 31.62 & 0.9513 & 0.6865 & 0.00 & 3.740e+56 & 0.3581 & 0.3960 \\
n144 & 11717.8 & 31.60 & 0.9513 & 0.6866 & 0.00 & 3.760e+56 & 0.3585 & 0.3964 \\
$n_\infty$ & 11701.1 & 31.58 & 0.9513 & 0.6869 & 0.00 & 3.778e+56 & 0.3586 & 0.3966 \\
\hline
$n$ & 3.65 & 5.45 & - - 
& 2.72 & 0.00 & 4.11 & 5.97 & 5.55 \\
\hline 
\end{tabular}
\end{table*}

\begin{table*}
\caption{Convergence of number of orbits $N$, merger time $t_{\rm{m}}/M$, remnant mass $M_f$, spin $\chi_f$ and recoil $V_f$, 
peak luminosity $\mathcal{L}_{\rm{peak}}$,  peak frequency $M\omega_{22}^{\rm{peak}}$, and
strain peak amplitude $\left(r/M\right)|h_{22}^{\rm{peak}}|$ for the RIT:BBH:1283 configuration with initial separation $D=24.64 M$, $\chi_{1,2} = 0.0$, $q=1.0$ and $\mathrm{f}=0.15$. Richardson extrapolation is used to determine convergence order $n$ and extrapolated values at $n_\infty$. 
}
\label{tab:convergence_Jim2}

\centering
\begin{tabular}{cccccccccc}
\hline
resolution  & $t_{\rm{m}}/M$ & $N$ & $M_f/M$ & $\chi_f$ & $V_f\,$[km/s] & $\mathcal{L}_{\rm{peak}}\,$[erg/s] 
  &$M\omega_{22}^{\rm{peak}}$ &$\left(r/M\right)|h_{22}^{\rm{peak}}|$ \\
\hline
n100 & 6524.0 & 20.93 & 0.9512 & 0.6851 & 0.00 & 3.657e+56 & 0.3568 & 0.3934 \\
n120 & 6526.9 & 20.94 & 0.9512 & 0.6850 & 0.00 & 3.685e+56 & 0.3572 & 0.3936 \\
n144 & 6527.8 & 20.94 & 0.9512 & 0.6850 & 0.00 & 3.670e+56 & 0.3576 & 0.3936 \\
$n_\infty$ & 6528.1 & 20.94 & 0.9512
& 0.6850 & 0.00 & 3.716e+56 & 0.3580 & 0.3937 \\
\hline
$n$ & 6.92 & 5.13 & - -
& 13.28 & 0.00 & 3.44 & - -
& 3.01 \\
\hline 
\end{tabular}
\end{table*}

\begin{table*}
  \caption{Convergence of number of orbits $N$, merger time $t_{\rm{m}}/M$ for this
    near plunge configuration, remnant mass $M_f$, spin $\chi_f$ and recoil $V_f$, 
peak luminosity $\mathcal{L}_{\rm{peak}}$,  peak frequency $M\omega_{22}^{\rm{peak}}$, and
strain peak amplitude $\left(r/M\right)|h_{22}^{\rm{peak}}|$ for the RIT:BBH:1637 configuration with initial separation $D=24.62 M$, $\chi_{1,2}^x = 0.70$, $q=1.0$ and $\mathrm{f}=0.40$. Richardson extrapolation is used to determine convergence order $n$ and extrapolated values at $n_\infty$. 
}
\label{tab:convergence_Jim3}

\centering
\begin{tabular}{cccccccccc}
\hline
resolution  & $t_{\rm{m}}/M$ & $N$ & $M_f/M$ & $\chi_f$ & $V_f\,$[km/s] & $\mathcal{L}_{\rm{peak}}\,$[erg/s] 
  &$M\omega_{22}^{\rm{peak}}$ &$\left(r/M\right)|h_{22}^{\rm{peak}}|$ \\
\hline
n100 & 228.0 & 1.04 & 0.9578 & 0.7833 & 0.00 & 4.952e+56 & 0.3695 & 0.3866 \\
n120 & 227.8 & 1.03 & 0.9578 & 0.7833 & 0.00 & 5.024e+56 & 0.3768 & 0.3892 \\
n144 & 228.0 & 1.04 & 0.9578 & 0.7833 & 0.00 & 5.062e+56 & 0.3717 & 0.3884 \\
$n_\infty$ & 228.0 & 1.04  & 0.9578 & 0.7833 & 0.00 & 5.103e+56 & 0.3593 & 0.3880 \\
\hline
$n$ & - - & - -
& 2.22 & - -
& 0.00 & 3.55 & 1.91 & 6.56 \\
\hline 
\end{tabular}
\end{table*}

\begin{table*}
  \caption{Convergence of number of orbits $N$, merger time $t_{\rm{m}}/M$ for a near
    plunge configuration, remnant mass $M_f$, spin $\chi_f$ and recoil $V_f$, 
peak luminosity $\mathcal{L}_{\rm{peak}}$,  peak frequency $M\omega_{22}^{\rm{peak}}$, and
strain peak amplitude $\left(r/M\right)|h_{22}^{\rm{peak}}|$ for the RIT:BBH:1670 configuration with initial separation $D=24.64 M$, $\chi_{1}^x = 0.70$, $\chi_{2}^y = 0.70$, $q=1.0$ and $\mathrm{f}=0.40$. Richardson extrapolation is used to determine convergence order $n$ and extrapolated values at $n_\infty$. 
}
\label{tab:convergence_Jim4}

\centering
\begin{tabular}{cccccccccc}
\hline
resolution  & $t_{\rm{m}}/M$ & $N$ & $M_f/M$ & $\chi_f$ & $V_f\,$[km/s] & $\mathcal{L}_{\rm{peak}}\,$[erg/s] 
  &$M\omega_{22}^{\rm{peak}}$ &$\left(r/M\right)|h_{22}^{\rm{peak}}|$ \\
\hline
n100 & 226.67 & 0.99 & 0.9623 & 0.75126 & 2052.7 & 4.339e+56 & 0.3649 & 0.4752 \\
n120 & 226.67 & 0.99 & 0.9623 & 0.75126 & 2080.3 & 4.381e+56 & 0.3711 & 0.4733 \\
n144 & 226.62 & 1.0 & 0.9623 & 0.75125 & 2078.6 & 4.405e+56 & 0.3706 & 0.4742 \\
$n_\infty$ 
& 226.62 & 1.0 & 0.9623 & 0.75125
& 2078.5 & 4.438e+56 & 0.3706 & 0.4749 \\
\hline
$n$ & - -
& - - & - - & - -
& 15.25 & 2.99 & 14.52 & 4.13 \\
\hline 
\end{tabular}
\end{table*}

The next round of convergence studies focus on long term evolutions equal mass, eccentric nonspinning as
in Table~\ref{tab:convergence_eBBH_q1} with 18 orbits to merger and a quasicircular, spinning with
$\chi_i^z=-1/3$ as given in Table~\ref{tab:convergence_eBBHspin_q1_shu-0.25} taking 23 orbits to merger. Those are all new simulations corresponding to the fifth release and labeled as RIT:BBH:2013 and RIT:BBH:2104 respectively.

\begin{table*}
  \caption{Convergence of merger time $t_{\rm{m}}/M$, number of orbits $N$, remnant mass $M_f$ and spin $\chi_f$, strain peak amplitude $\left(r/M\right)|h_{22}^{\rm{peak}}|$, peak frequency $M\omega_{22}^{\rm{peak}}$  and peak luminosity $\mathcal{L}_{\rm{peak}}$ for eBBH:03 from \cite{Ficarra:2024nro} and relabeled as RIT:BBH:2013 in the RIT Catalog with spinless $q=1.0$ BBH starting at $D/M=19.88$, $\chi_{1,2} = 0.0$, $q=1.0$ and $\mathrm{f}=0.10$.
Richardson extrapolation is used to determine convergence order $n$ and extrapolated values at $n_\infty$.
\label{tab:convergence_eBBH_q1}
}
\centering
\begin{tabular}{cccccccc}
\hline
resolution & $t_{\rm{m}}/M$ & $N$ & $M_f/M$ & $\chi_f$ & $\left(r/M\right)|h_{22}^{\rm{peak}}|$ & $M\omega_{22}^{\rm{peak}}$ & $\mathcal{L}_{\rm{peak}}\,$[erg/s]\\
\hline
$n$100 & 5186.7 & 18.93 & 0.9517 & 0.6857 & 0.3915 & 0.3570 & 3.603e+56 \\
$n$120 & 4787.6 & 17.77 & 0.9520 & 0.6868 & 0.3930 & 0.3582 & 3.663e+56 \\
$n$144 & 4771.6 & 17.73 & 0.9520 & 0.6864 & 0.3927 & 0.3585 & 3.678e+56 \\
$n_\infty$ & 4770.9 & 17.72 & 0.9520 & 0.6865 & 0.3928 & 0.3586 & 3.683e+56 \\
\hline
$n$ & 17.62 & 17.71 & 15.07 & 5.99 & 8.54 & 7.86 & 7.74 \\
\hline 
\end{tabular}
\end{table*}

\begin{table*}
\caption{Convergence of merger time $t_{\rm{m}}/M$, number of orbits $N$, remnant mass $M_f$ and spin $\chi_f$, strain peak amplitude $\left(r/M\right)|h_{22}^{\rm{peak}}|$, peak frequency $M\omega_{22}^{\rm{peak}}$ and peak luminosity $\mathcal{L}_{\rm{peak}}$ for the RIT:BBH:2104 configuration with initial separation $D= 17.10M$, $\chi_{hu} = -0.25$, $q=1$ and $\mathrm{f}=0.00$ . Richardson extrapolation is used to determine convergence order $n$ and extrapolated values at $n_\infty$.
\label{tab:convergence_eBBHspin_q1_shu-0.25}
}
\centering
\begin{tabular}{cccccccc}
\hline
resolution & $t_{\rm{m}}/M$ & $N$ & $M_f/M$ & $\chi_f$ & $\left(r/M\right)|h_{22}^{\rm{peak}}|$ & $M\omega_{22}^{\rm{peak}}$ & $\mathcal{L}_{\rm{peak}}\,$[erg/s] \\
\hline
$n$100 & 7471.8 & 23.36 & 0.9591 & 0.5791 & 0.3942 & 0.3352 & 3.173e+56 \\
$n$120 & 7235.6 & 22.78 & 0.9591 & 0.5806 & 0.3945 & 0.3362 & 3.204e+56 \\
$n$144 & 7150.0 & 22.58 & 0.9592 & 0.5812 & 0.3946 & 0.3364 & 3.224e+56 \\
$n_\infty$ & 7101.3 & 22.46 & 0.9592 & 0.5816 & 0.3948 & 0.3364 & 3.257e+56\\
\hline
$n$ & 5.56 & 5.67 & 5.22 & 4.82 & 2.90 & 11.81 & 2.52 \\
\hline 
\end{tabular}
\end{table*}


The other two set of new simulations correspond to equal mass, highly eccentric $\mathrm{f}=0.20$ and highly
spinning with $\chi_i^z=+2/3$ as given in Table~\ref{tab:convergence_eBBHspin_q1_shu0.5_f0.2}
leading to 22 orbits to merger in the case of RIT:BBH:2153, while in Table~\ref{tab:convergence_eBBHspin_q1_shu-0.5_f0.2} we give the results of reversing the spins to $\chi_i^z=-2/3$ leading to
20 orbits to merger for RIT:BBH:2170.

\begin{table*}
\caption{Convergence of merger time $t_{\rm{m}}/M$, number of orbits $N$, remnant mass $M_f$ and spin $\chi_f$, strain peak amplitude $\left(r/M\right)|h_{22}^{\rm{peak}}|$, peak frequency $M\omega_{22}^{\rm{peak}}$ and peak luminosity $\mathcal{L}_{\rm{peak}}$ for the RIT:BBH:2153 configuration with initial separation $D=25.84 M$, $\chi_{hu} = 0.5$, $q=1$ and $\mathrm{f}=0.20$. Richardson extrapolation is used to determine convergence order $n$ and extrapolated values at $n_\infty$.
\label{tab:convergence_eBBHspin_q1_shu0.5_f0.2}
}
\centering
\begin{tabular}{cccccccc}
\hline
resolution & $t_{\rm{m}}/M$ & $N$ & $M_f/M$ & $\chi_f$ & $\left(r/M\right)|h_{22}^{\rm{peak}}|$ & $M\omega_{22}^{\rm{peak}}$ & $\mathcal{L}_{\rm{peak}}\,$[erg/s] \\
\hline
$n$100 & 6117.8 & 21.80 & 0.9222 & 0.8771 & 0.3922 & 0.4237 & 5.24e+56 \\
$n$120 & 6091.9 & 21.73 & 0.9226 & 0.8753 & 0.3920 & 0.4211 & 5.28e+56 \\
$n$144 & 6086.9 & 21.74 & 0.9227 & 0.8747 & 0.3917 & 0.4211 & 5.33e+56 \\
$n_\infty$ & 6085.7 & 21.73 & 0.9229 & 0.8744 & 0.3915 & 0.4211 & {5.37e+56} \\
\hline
$n$ & 9.02 & 13.36 & 4.45 & 5.53 & - -
& 26.95 & - - \\
\hline 
\end{tabular}
\end{table*}

\begin{table*}
\caption{Convergence of merger time $t_{\rm{m}}/M$, number of orbits $N$, remnant mass $M_f$ and spin $\chi_f$, strain peak amplitude $\left(r/M\right)|h_{22}^{\rm{peak}}|$, peak frequency $M\omega_{22}^{\rm{peak}}$ and peak luminosity $\mathcal{L}_{\rm{peak}}$ for the RIT:BBH:2170 configuration with initial separation $D=31.58 M$, $\chi_{hu} = -0.5$, $q=1$ and $\mathrm{f}=0.20$. Richardson extrapolation is used to determine convergence order $n$ and extrapolated values at $n_\infty$.
\label{tab:convergence_eBBHspin_q1_shu-0.5_f0.2}
}
\centering
\begin{tabular}{cccccccc}
\hline
resolution & $t_{\rm{m}}/M$ & $N$ & $M_f/M$ & $\chi_f$ & $\left(r/M\right)|h_{22}^{\rm{peak}}|$ & $M\omega_{22}^{\rm{peak}}$ & $\mathcal{L}_{\rm{peak}}\,$[erg/s] \\
\hline
$n$100 & 8116.2 & 21.03 & 0.9647 & 0.467 & 0.3970 & 0.3151 & 2.85e+56 \\
$n$120 & 7839.8 & 20.34 & 0.9642 & 0.468 & 0.3939 & 0.3151 & 2.82e+56 \\
$n$144 & 7767.2 & 20.14 & 0.9639 & 0.471 & 0.3964 & 0.3142 & 2.89e+56 \\
$n_\infty$ & 7741.4 & 20.05 & 0.9637 & 0.472 & 0.3956 & 0.3140 & 2.90e+56 \\
\hline
$n$ & 7.34 & 6.74 & 4.36 & {- -} & 1.23 & {- -} & {- -} \\
\hline 
\end{tabular}
\end{table*}


The other two cases of interest involve unequal mass ratios still with large eccentricities $\mathrm{f}=0.20$
and large negative spins. In Table~\ref{tab:convergence_eBBHspin_q2} we report the convergence
properties of a $q\approx1/2$ mass ratio binary with $\chi_i^z=-0.64$ labeled as RIT:BBH:2172 and
leading to 25 orbits to merger simulation. While Table~\ref{tab:convergence_eBBHspin_q8_shu-0.5}
gives the convergence study of a more unequal mass case, $q\approx1/8.5$ and spins $\chi_i^z=-0.55$
corresponding to the catalog run RIT:BBH:2146 that leads to a simulation that takes 33 orbits
to merger.

\begin{table*}
\caption{Convergence of merger time $t_{\rm{m}}/M$, number of orbits $N$, remnant mass $M_f$, spin $\chi_f$ and recoil $V_f$, strain peak amplitude $\left(r/M\right)|h_{22}^{\rm{peak}}|$, peak frequency $M\omega_{22}^{\rm{peak}}$ and peak luminosity $\mathcal{L}_{\rm{peak}}$ for the RIT:BBH:2172 configuration with initial separation $D=32.83 M$, $\chi_{hu} = -0.5$, $q=0.4776$ and $\mathrm{f}=0.20$. Richardson extrapolation is used to determine convergence order $n$ and extrapolated values at $n_\infty$. 
\label{tab:convergence_eBBHspin_q2}
}
\centering
\begin{tabular}{cccccccccc}
\hline
resolution & $t_{\rm{m}}/M$ & $N$ & $M_f/M$ & $\chi_f$ & $V_f\,$[km/s] & $\left(r/M\right)|h_{22}^{\rm{peak}}|$ & $M\omega_{22}^{\rm{peak}}$ & $\mathcal{L}_{\rm{peak}}\,$[erg/s] \\
\hline
$n$100 & 9107.8 & 22.66 & 0.9715 & 0.3821 & 245.37 & 0.3470 & 0.32027 & 2.207e+56 \\
$n$120 & 10134.4 & 24.92 & 0.9724 & 0.3674 & 251.52 & 0.3374 & 0.2866 & 1.972e+56 \\
$n$144 & 10548.9 & 25.56 & 0.9719 & 0.3613 & 251.23 & 0.3399 & 0.3057 & 2.097e+56 \\
$n_\infty$ & 10829.6 & 25.82 & 0.9721 & 0.3572 & 251.24 & 0.3394 & 0.2988 & 2.054e+56\\
\hline
$n$ & 4.97 & 6.89 & 3.00 & 4.91 & 16.75 & 7.33 & 3.10 & 3.47 \\
\hline 
\end{tabular}
\end{table*}

\begin{table*}
\caption{Convergence of merger time $t_{\rm{m}}/M$, number of orbits $N$, remnant mass $M_f$, spin $\chi_f$ and recoil $V_f$, strain peak amplitude $\left(r/M\right)|h_{22}^{\rm{peak}}|$ , peak frequency $M\omega_{22}^{\rm{peak}}$ and peak luminosity $\mathcal{L}_{\rm{peak}}$ for the RIT:BBH:2146 configuration with initial separation $D=24.48 M$, $\chi_{hu} = -0.5$, $q=0.1170$ and $\mathrm{f}=0.20$. Richardson extrapolation is used to determine convergence order $n$ and extrapolated values at $n_\infty$.
\label{tab:convergence_eBBHspin_q8_shu-0.5}
}
\centering
\begin{tabular}{cccccccccc}
\hline
resolution & $t_{\rm{m}}/M$ & $N$ & $M_f/M$ & $\chi_f$ & $V_f\,$[km/s] & $\left(r/M\right)|h_{22}^{\rm{peak}}|$ & $M\omega_{22}^{\rm{peak}}$ & $\mathcal{L}_{\rm{peak}}\,$[erg/s] \\
\hline
$n$100 & 2289.2 & 8.47  & 0.993 & -0.1081 & 109.4 & 0.1411 & 0.2671 & 3.566e+55 \\
$n$120 & 3032.3 & 12.08 & 0.993 & -0.1171 & 99.52 & 0.1322 & 0.2593 & 3.198e+55 \\
$n$144 & 3282.8 & 12.47 & 0.992 & -0.1143 & 103.22 & 0.1363 & 0.2615 & 3.343e+55 \\
$n_\infty$ & 3410.3 & 12.51 & 0.991 & -0.1150 & 102.21 & 0.1350 & 0.2610 & 3.302e+55 \\
\hline
$n$ & 5.97 & 12.25 & {- -} & 6.45 & 5.39 & 4.28 & 7.04 & 5.11 \\
\hline 
\end{tabular}
\end{table*}

\subsection{A six resolutions case study}\label{sec:6convergence}

This configuration eBBH:08 from \cite{Ficarra:2024nro} and relabeled as RIT:BBH:2010 in this release of
the RIT Catalog, for spinless $q=0.4776$ BBH starting at $D/M=19.26$ was of particular interest since
it represented our best match \cite{McMillin:2025hof} to the LVK gravitational wave event GW200208\_22.
We note that the n144 case has an error fluctuation already seen in the LnL evaluations,
hence will reduce the convergence order when included in the triad of resolutions.

In order to further study the convergence properties of the eccentric simulations we have considered an extended
range of global grid resolutions. To the standard n100-n120-n144 three resolutions studied above, we have
considered a lowest resolution n084 and two higher n172 and n208 resolutions. The inclusion of the n084,
that we expect to be below the convergence regime, will give us some insight on how this is approached
from below, while the n172, n208 might give us how over-resolving the problems behaves.

Table~\ref{tab:convergence_eBBH_q2} we display the result of the 6 simulations, its convergence rates,
and extrapolations to infinite resolutions. We first note that while the inclusion of n084 into the
radiative quantities such as the recoil velocity and waveform peaks lead to not a clean convergence
rate, the values themselves do not fall much apart from that of the much higher resolution runs.
We have also profited from the combinations with n120 and n172 to square the split factor $f^2=1.2^2=1.44$,
what leads to good convergence rates for those quantities due to further split in their values.
The same beneficial effect is observed in the combination n100-n144-n208.

We also observe the greater robustness of the convergence order when we analyze the horizon quantities
as the final hole mass and spin as well as the track properties time and number of orbits to merger.
The values of the horizon quantities are particularly tight together and difficult to measure differences
beyond 4 digits. In particular the final horizon mass benefits of the larger splitting in resolutions
provided by the n100-n144-n208 combination.

\begin{table*}
  \caption{Convergence of merger time $t_{\rm{m}}/M$, number of orbits $N$, remnant mass $M_f$, spin $\chi_f$ and recoil $V_f$, strain peak amplitude $\left(r/M\right)|h_{22}^{\rm{peak}}|$, peak frequency $M\omega_{22}^{\rm{peak}}$ and peak luminosity $\mathcal{L}_{\rm{peak}}$ for eBBH:08 from \cite{Ficarra:2024nro} and relabeled as RIT:BBH:2010 in the RIT Catalog with spinless $q=0.4776$ BBH starting at $D/M=19.26$, $\chi_{1,2} = 0.0$, $q=1.0$ and $\mathrm{f}=0.10$.
Richardson extrapolation is used to determine convergence order $n$ and extrapolated values at $n_\infty$.
\label{tab:convergence_eBBH_q2}
}
\centering
\begin{tabular}{c|cccc|cccc}
\hline
resolution & $t_{\rm{m}}/M$ & $N$ & $M_f/M$ & $\chi_f$ & $V_f\,$[km/s] & $\left(r/M\right)|h_{22}^{\rm{peak}}|$ & $M\omega_{22}^{\rm{peak}}$ & $\mathcal{L}_{\rm{peak}}\,$[erg/s] \\
\hline
$n$084 & 4908.7 & 18.66 & 0.9618 & 0.6156 & 148.81 & 0.3374 & 0.3408 & 2.664e+56 \\ 
$n$100 & 4611.1 & 17.82 & 0.9625 & 0.6140 & 152.38 & 0.3353 & 0.3427 & 2.629e+56 \\
$n$120 & 4649.6 & 17.95 & 0.9626 & 0.6147 & 156.72 & 0.3371 & 0.3451 & 2.676e+56 \\
$n$144 & 4629.5 & 17.88 & 0.9626 & 0.6143 & 158.81 & 0.3372 & 0.3445 & 2.678e+56 \\
$n$172 & 4643.2 & 17.93 & 0.9626 & 0.6146 & 159.97 & 0.3369 & 0.3447 & 2.689e+56 \\
$n$208 & 4645.3 & 17.94 & 0.9626 & 0.6146 & 160.51 & 0.3373 & 0.3450 & 2.692e+56 \\
\hline
$n_\infty$(084-100-120) & 4645.2 & 17.93 & 0.9627 & 0.61452 & - - & - - & - - & - - \\ 
$n_\infty$(100-120-144) & 4636.4 & 17.90 & 0.9626 & 0.61448 & 160.75 & 0.3372 & 0.3446 & 2.68e+56 \\
$n_\infty$(120-144-172) & 4637.7 & 17.91 & 0.9626 & 0.61449 & 161.42 & 0.3371 & 0.3446 & 2.68e+56\\ 
$n_\infty$(144-172-208) & 4645.6 & 17.94 & 0.9626 & 0.61464 & 160.98 & 0.3371 & 0.3442 & 2.69e+56 \\ 
\hline
$n$(084-100-120) & 11.22 & 10.23 & 8.57 & 4.24 & - - & - - & - - & - - \\ 
$n$(100-120-144) & 3.56 & 3.58 & 6.76 & 3.00 & 4.01 & 15.33 & 7.75 & 19.02 \\
$n$(120-144-172) & 2.10 & 1.85 & - - & 2.51 & 3.23 & - - & 6.53 & - - \\ 
$n$(144-172-208) & 10.44 & 8.83 & 12.90 & 10.53 & 4.19 & - - & - - & 8.09 \\ 
\hline
$n_\infty$(084-120-172) & 4643.1 & 17.93 & 0.9626 & 0.6146 & 162.23 & 0.3368 & 0.3447 & 2.69e+56 \\
$n_\infty$(100-144-208) & 4645.0 & 17.94 & 0.9627 & 0.6146 & 161.12 & 0.3373 & 0.3451 & 2.70e+56\\
\hline
$n$(084-120-172) & 10.15 & 9.79 & - - & 4.67 & 2.44 & 1.88 & 6.44 & - - \\
$n$(100-144-208) & - - & - - & 2.12 & - - & 3.65 & 7.17 & 3.76 & 3.38 \\
\hline 
\end{tabular}
\end{table*}


It is interesting to compare some of this results with similar configurations for quasicircular
orbits. For instance from the results in \cite{Healy:2017mvh}, we observe that the recoil velocity
for $q=1/2$ is about 155km/s, $M\omega_{22}^{\rm{peak}}=0.3463$, $\left(r/M\right)|h_{22}^{\rm{peak}}|=0.3451$, and
$\mathcal{L}_{\rm{peak}}\,=2.82e+56$[erg/s], well in line with the results we are finding.

\section{Tables of initial data and results of the new simulations}\label{app:ID}

In this appendix we provide tables with the relevant BBH
configuration details.  In Table \ref{tab:qcID}, we provide the
initial data parameters for the new 51 quasicircular configurations
used to start the full numerical evolutions. Those include 1 nonspinning,
13 spin aligned (nonprecessing) and 37 misaligned spins (precessing) binaries.
In Table \ref{tab:eccID}, we provide the
initial data parameters for the new 65 nonspinning and 132 aligned spinning
eccentric configurations used to start the full numerical evolutions
for the apastron by use of the prescription of
decreasing the quasicircular
tangential orbital momentum by a factor (1-f).


In Tables \ref{tab:IDr} and \ref{tab:IDr_prec},
we provide the binary mass and spin parameters after 
they settle into a more physical value after radiating and absorbing
the spurious gravitation wave content from the initial mathematical 
choice of conformal flatness.  These relaxed values are calculated at  
a fiducial $t=200m$.  Table \ref{tab:IDr} includes the new 14 quasicircular
simulations (1 nonspinning) and the new 197 (65 nonspinning) eccentric
binaries.
Table \ref{tab:IDr_prec} adds the new 37 precessing (plus 4 higher resolution
studies of release 4 runs)
quasicircular black hole binaries evolutions.

Finally, In Table \ref{tab:spinerad}, we give
the values of the gravitational power and linear momentum
radiated during evolution and the peak amplitude and
frequency of the strain (2,2) mode as well as
the final black hole mass and spin as measured through the (accurate)
isolated horizon formalism \cite{Dreyer02a,Campanelli:2006fy} for all the new simulations
reported in this paper.


\input{id_table.tex}

\input{relaxed_table.tex}

\input{final_state_table.tex}

\end{document}

%% file: id_table.tex
\begin{longtable*}{lccccccccccccc}
\caption{Initial data parameters for the quasi-circular
configurations with a smaller mass black hole (labeled 1),
and a larger mass spinning black hole (labeled 2). The punctures are located
at $\vec r_1 = (x_1,0,0)$ and $\vec r_2 = (x_2,0,0)$, with individual linear momenta
$P=\pm (P_r, P_t,0)$, spin magnitudes $|S_i|$, puncture mass parameters
$m^p/m$, horizon (Christodoulou) masses $m^H/m$, total ADM mass
$M_{\rm ADM}$, and dimensionless spins $|a/m_H| = |S/m_H^2|$.
The spin directions for the precessing simulations are given in
the catalog and in Tab.~\ref{tab:IDr_prec}.}
\label{tab:qcID}\\
\hline
\hline
Run   & $x_1/m$ & $x_2/m$  & $P_r/m$    & $P_t/m$ & $m^p_1/m$ & $m^p_2/m$ & $|S_1/m^2|$ & $|S_2/m^2|$ & $m^H_1/m$ & $m^H_2/m$ & $M_{\rm ADM}/m$ & $|a_1/m_1^H|$ & $|a_2/m_2^H|$\\
\hline
\endfirsthead

\multicolumn{14}{c}
{{\tablename\ \thetable{} -- continued from previous
page}}\\
\hline
Run   & $x_1/m$ & $x_2/m$  & $P_r/m$    & $P_t/m$ & $m^p_1/m$ & $m^p_2/m$ & $|S_1/m^2|$ & $|S_2/m^2|$ & $m^H_1/m$ & $m^H_2/m$ & $M_{\rm ADM}/m$ & $|a_1/m_1^H|$ & $|a_2/m_2^H|$\\
\hline
\endhead
\hline \multicolumn{14}{r}{{Continued on next page}} \\
\hline
\endfoot
\hline\hline
\endlastfoot

RIT:BBH:1924 n100 & -10.12 & 1.67 & -1.18e-04 & 0.04068 & 0.1367 & 0.53 & 0 & 0.5878 & 0.1429 & 0.8571 & 0.9955 & 0 & 0.8 \\
RIT:BBH:1925 n100 & -10.12 & 1.67 & -1.18e-04 & 0.04068 & 0.1367 & 0.53 & 0 & 0.5878 & 0.1429 & 0.8571 & 0.9955 & 0 & 0.8 \\
RIT:BBH:1926 n100 & -10.12 & 1.67 & -1.18e-04 & 0.0407 & 0.1367 & 0.5299 & 0 & 0.5878 & 0.1429 & 0.8571 & 0.9955 & 0 & 0.8 \\
RIT:BBH:1927 n100 & -10.13 & 1.67 & -1.19e-04 & 0.04072 & 0.1367 & 0.5299 & 0 & 0.5878 & 0.1429 & 0.8571 & 0.9955 & 0 & 0.8 \\
RIT:BBH:1928 n100 & -10.12 & 1.67 & -1.18e-04 & 0.0407 & 0.1367 & 0.5299 & 0 & 0.5878 & 0.1429 & 0.8571 & 0.9955 & 0 & 0.8 \\
RIT:BBH:1929 n100 & -10.13 & 1.67 & -1.19e-04 & 0.04073 & 0.1367 & 0.5299 & 0 & 0.5878 & 0.1429 & 0.8571 & 0.9955 & 0 & 0.8 \\
RIT:BBH:1930 n100 & -10.13 & 1.68 & -1.23e-04 & 0.04118 & 0.1367 & 0.5299 & 0 & 0.5878 & 0.1429 & 0.8571 & 0.9955 & 0 & 0.8 \\
RIT:BBH:1931 n100 & -10.13 & 1.68 & -1.23e-04 & 0.04119 & 0.1367 & 0.5299 & 0 & 0.5878 & 0.1429 & 0.8571 & 0.9955 & 0 & 0.8 \\
\end{longtable*}

\begin{longtable*}{lccccccccccccc}
  \caption{Initial data parameters for the new nonspinning and aligned-spin
    eccentric configurations with a smaller mass black hole (labeled 1),
and a larger mass spinning black hole (labeled 2). The punctures are located
at $\vec r_1 = (x_1,0,0)$ and $\vec r_2 = (x_2,0,0)$, with individual linear momenta
$P=\pm (P_r, P_t,0)$, spin magnitudes $|S_i|$, puncture mass parameters
$m^p/m$, horizon (Christodoulou) masses $m^H/m$, total ADM mass
$M_{\rm ADM}$ and initial post-Newtonian eccentricity. Also given is the 
number of orbits to merger, $N_{orb}$, and dimensionless spins $|a/m_H| = |S/m_H^2|$. }
\label{tab:eccID}\\
\hline
\hline
Run   & $x_1/m$ & $x_2/m$  & $P_r/m$    & $P_t/m$ & $m^p_1/m$ & $m^p_2/m$ & $m^H_1/m$ & $m^H_2/m$ & $M_{\rm ADM}/m$ & $e$ & $N_{orb}$ & $|a_1/m_1^H|$ & $|a_2/m_2^H|$\\
\hline
\endfirsthead

\multicolumn{14}{c}
{{\tablename\ \thetable{} -- continued from previous
page}}\\
\hline
Run   & $x_1/m$ & $x_2/m$  & $P_r/m$    & $P_t/m$ & $m^p_1/m$ & $m^p_2/m$ & $m^H_1/m$ & $m^H_2/m$ & $M_{\rm ADM}/m$ & $e$ & $N_{orb}$ & $|a_1/m_1^H|$ & $|a_2/m_2^H|$\\
\hline
\endhead
\hline \multicolumn{14}{r}{{Continued on next page}} \\
\hline
\endfoot
\hline\hline
\endlastfoot

RIT:BBH:1282 n120 & -12.32 & 12.32 & 0 & 0.0491 & 0.4944 & 0.4944 & 0.5 & 0.5 & 0.9943 & 0.2314 & 31.62 & 0 & 0 \\
RIT:BBH:1282 n144 & -12.32 & 12.32 & 0 & 0.0491 & 0.4944 & 0.4944 & 0.5 & 0.5 & 0.9943 & 0.2314 & 31.60 & 0 & 0 \\
RIT:BBH:1283 n120 & -12.32 & 12.32 & 0 & 0.04637 & 0.4945 & 0.4945 & 0.5 & 0.5 & 0.9938 & 0.3419 & 20.94 & 0 & 0 \\
RIT:BBH:1283 n144 & -12.32 & 12.32 & 0 & 0.04637 & 0.4945 & 0.4945 & 0.5 & 0.5 & 0.9938 & 0.3419 & 20.94 & 0 & 0 \\
RIT:BBH:1637 n120 & -12.31 & 12.31 & 0 & 0.03274 & 0.3693 & 0.3693 & 0.5 & 0.5 & 0.992 & 0.7645 & 1.03 & 0.7 & 0.7 \\
RIT:BBH:1637 n144 & -12.31 & 12.31 & 0 & 0.03274 & 0.3693 & 0.3693 & 0.5 & 0.5 & 0.992 & 0.7645 & 1.04 & 0.7 & 0.7 \\
RIT:BBH:1670 n120 & -12.32 & 12.32 & 0 & 0.03273 & 0.3693 & 0.3693 & 0.5 & 0.5 & 0.992 & 0.7677 & 1.23 & 0.7 & 0.7 \\
RIT:BBH:1670 n144 & -12.32 & 12.32 & 0 & 0.03273 & 0.3693 & 0.3693 & 0.5 & 0.5 & 0.992 & 0.7677 & 1.23 & 0.7 & 0.7 \\
\end{longtable*}

%% file: relaxed_table.tex
\clearpage
\begin{longtable*}{cccccc}
\caption{The masses and spins of the nonprecessing BHBs in Tables~\ref{tab:qcID} and \ref{tab:eccID} after the
BHs had time to equilibrate ($t/m=200$).  
}
\label{tab:IDr}\\
\hline
\hline
Run & $q^r$ & $m^r_1/m$ & $m^r_2/m$ & $\chi^r_{1z}$ & $\chi^r_{2z}$  \\
\hline
\endfirsthead

\multicolumn{6}{c}
{{\tablename\ \thetable{} -- continued from previous page}}\\
\hline
Run & $q^r$ & $m^r_1/m$ & $m^r_2/m$ & $\chi^r_{1z}$ & $\chi^r_{2z}$  \\
\hline
\endhead
\hline \multicolumn{6}{r}{{Continued on next page}} \\
\hline
\endfoot
\hline\hline
\endlastfoot

RIT:BBH:1936 n144 &	$0.2502$ & $0.2000$ & $0.7996$ & $-0.9499$ & $-0.9495$ \\
RIT:BBH:1942 n144 &	$0.1426$ & $0.1247$ & $0.8745$ & $-0.9493$ & $-0.9495$ \\
RIT:BBH:1951 n144 &	$0.2502$ & $0.2000$ & $0.7996$ & $0.9499$ & $-0.9495$ \\
RIT:BBH:1952 n144 &	$0.1426$ & $0.1247$ & $0.8744$ & $0.9493$ & $-0.9494$ \\
RIT:BBH:1959 n144 &	$0.2502$ & $0.2000$ & $0.7996$ & $0.9499$ & $0.9496$ \\
RIT:BBH:1960 n144 &	$0.2502$ & $0.2000$ & $0.7996$ & $-0.9499$ & $0.9496$ \\
RIT:BBH:1961 n144 &	$0.1427$ & $0.1248$ & $0.8745$ & $-0.9495$ & $0.9495$ \\
RIT:BBH:1962 n144 &	$0.1427$ & $0.1248$ & $0.8745$ & $0.9496$ & $0.9495$ \\
\end{longtable*}

\clearpage
\begin{longtable*}{lcccrrrrrr}
\caption{The mass and spin of the QC precessing BHBs in Table~\ref{tab:qcID} after the
BHs had time to equilibrate ($t/m=200$).}
\label{tab:IDr_prec}\\
\hline
\hline
Run & $q^r$ & $m^r_1/m$ & $m^r_2/m$ & $\chi^r_{1x}$ & $\chi^r_{1y}$ &  $\chi^r_{1z}$ &  $\chi^r_{2x}$ &  $\chi^r_{2y}$ & $\chi^r_{2z}$ \\
\hline
\endfirsthead

\multicolumn{10}{c}
{{\tablename\ \thetable{} -- continued from previous page}}\\
\hline
Run & $q^r$ & $m^r_1/m$ & $m^r_2/m$ & $\chi^r_{1x}$ & $\chi^r_{1y}$ &  $\chi^r_{1z}$ &  $\chi^r_{2x}$ &  $\chi^r_{2y}$ & $\chi^r_{2z}$ \\
\hline
\endhead
\hline \multicolumn{10}{r}{{Continued on next page}} \\
\hline
\endfoot
\hline\hline
\endlastfoot

RIT:BBH:1924 n100 &	$0.1668$ & $0.1429$ & $0.8567$ & 0.0 & 0.0 & 0.0 & 0.3984 & 0.0295 & 0.6942 \\
RIT:BBH:1925 n100 &	$0.1668$ & $0.1429$ & $0.8567$ & 0.0 & 0.0 & 0.0 & -0.3774 & -0.1315 & 0.6941 \\
RIT:BBH:1926 n100 &	$0.1668$ & $0.1429$ & $0.8567$ & 0.0 & 0.0 & 0.0 & -0.3020 & 0.2609 & 0.6944 \\
RIT:BBH:1927 n100 &	$0.1668$ & $0.1429$ & $0.8567$ & 0.0 & 0.0 & 0.0 & 0.1308 & -0.3771 & 0.6944 \\
RIT:BBH:1928 n100 &	$0.1668$ & $0.1429$ & $0.8567$ & 0.0 & 0.0 & 0.0 & 0.2613 & 0.3025 & 0.6941 \\
RIT:BBH:1929 n100 &	$0.1668$ & $0.1429$ & $0.8567$ & 0.0 & 0.0 & 0.0 & -0.0287 & 0.3983 & 0.6943 \\
RIT:BBH:1930 n100 &	$0.1668$ & $0.1429$ & $0.8567$ & 0.0 & 0.0 & 0.0 & 0.6901 & 0.0565 & 0.4026 \\
RIT:BBH:1931 n100 &	$0.1668$ & $0.1429$ & $0.8567$ & 0.0 & 0.0 & 0.0 & -0.6522 & -0.2330 & 0.4023 \\
\end{longtable*}

%% file: final_state_table.tex
\clearpage
\begin{longtable*}{lcccccc}
\caption{The final state parameters of the system including the final mass, spin, recoil velocity, 
peak luminosity, and frequency and amplitude at the peak of the 22-strain mode, $M\omega^{\rm peak}_{22}$ and $(r/M)|h^{\rm peak}_{22}|$.
}
\label{tab:spinerad}\\
\hline
\hline
Run & $\mathcal{M}^{IH}$ & $\chi_{\mathrm{rem}}^{IH}$ & $V [km/s]$ & $L [ergs/s]$ & $M\omega^{\rm peak}_{22}$ & $(r/M)|h^{\rm peak}_{22}|$\\
\hline
\hline
\endfirsthead

\multicolumn{7}{c}
{{\tablename\ \thetable{} -- continued from previous page}}\\
\hline
Run & $\mathcal{M}^{IH}$ & $\chi_{\mathrm{rem}}^{IH}$ & $V [km/s]$ & $L [ergs/s]$ & $M\omega^{\rm peak}_{22}$ & $(r/M)|h^{\rm peak}_{22}|$\\
\hline
\endhead
\hline \multicolumn{7}{r}{{Continued on next page}} \\
\hline
\endfoot
\hline\hline
\endlastfoot

RIT:BBH:1924 n100 & 0.9711 & 0.8535 & 103.86 & 1.333e+56 & 0.3654 & 0.2000 \\
RIT:BBH:1925 n100 & 0.9712 & 0.8537 & 30.75 & 1.34e+56 & 0.48 & 0.1640 \\
RIT:BBH:1926 n100 & 0.9708 & 0.8528 & 531.40 & 1.488e+56 & 0.468 & 0.1998 \\
RIT:BBH:1927 n100 & 0.9712 & 0.8539 & 267.30 & 1.443e+56 & 0.2757 & 0.1673 \\
RIT:BBH:1928 n100 & 0.9712 & 0.8538 & 42.95 & 1.377e+56 & 0.372 & 0.1985 \\
RIT:BBH:1929 n100 & 0.9712 & 0.854 & 96.29 & 1.407e+56 & 0.3948 & 0.1986 \\
RIT:BBH:1930 n100 & 0.9769 & 0.8002 & 489.90 & 1.247e+56 & 0.1913 & 0.1305 \\
RIT:BBH:1931 n100 & 0.9769 & 0.8001 & 529.71 & 1.254e+56 & 0.3436 & 0.1573 \\
\end{longtable*}